\documentclass[a4paper,11pt]{article}
\pdfoutput=1 

\usepackage{jheppub} 

\usepackage[T1]{fontenc} 

\title{Vacuum decay and bubble nucleation in the anti-de Sitter black holes}

\author[a,b]{Ran Li,}
\author[b,c,*]{Jin Wang \note[*]{Corresponding author}}
\affiliation[a]{School of Physics, Henan Normal University, Xinxiang 453007, China}
\affiliation[b]{Department of Chemistry, Stony Brook University, Stony Brook, NY 11794, USA}
\affiliation[c]{Department of Physics and Astronomy, Stony Brook University, Stony Brook, NY 11794, USA,}

\emailAdd{liran@htu.edu.cn}
\emailAdd{jin.wang.1@stonybrook.edu}

\abstract{We study the vacuum decay and the bubble nucleation in the anti-de Sitter black holes. In the bubble nucleation spacetime, the interior and the exterior of the bubble wall are described by two anti-de Sitter black hole spacetimes with different cosmological constants. We calculate the Euclidean action of the bubble nucleation spacetime and give the numerical results of the tunneling rates for different cases. It is shown that the black hole can act as a source of inhomogeneities and catalyze the vacuum decay and the bubble nucleation in the anti-de Sitter spacetime. For the RNAdS black holes, the tunneling rate to the final RNAdS black hole with the minimum critical mass is the highest among all the possible tunneling channels.}

\begin{document} 
\maketitle
\flushbottom

\section{Introduction}
\label{sec:intro}

Black hole phase transition is a very interesting topic, which attracted much attention since Hawking's discovery of black hole thermodynamics \cite{Hawking:1975vcx}. One of the well known example of black hole phase transition is the Hawking-Page phase transition \cite{Hawking:1982dh}. That is there is a first order phase transition between the AdS black hole and the thermal AdS space. Inspired by the studies of the phase transition kinetics in condensed matter physics and polymer physics, the Schwarzschild AdS black hole was treated as a complex system with the horizon radius as the order parameter, based on which the thermodynamics and the kinetics of the Hawking-Page phase transition was investigated \cite{Li:2020khm}. In ordinary material system admitting first order phase transition, the bubble nucleation is the key mechanism to realize the phase transition process with spatial inhomogeneity \cite{Langer:1967}. The thermal activation of the bubble nucleation in Schwarzschild AdS black hole was investigated in \cite{Sasaki:2014spa,Chen:2017suz}. Interestingly, the bubble nucleation process that describes the complete evaporation of the Schwarzschild AdS black hole is found to have the same probability with that of the Hawking-Page phase transition from the black hole to the thermal radiation.

In recent years, the thermodynamics of charged AdS black hole in the extended phase space involving the variable cosmological constant, has been widely explored \cite{Kubiznak:2014zwa,Kubiznak:2016qmn}. The small/large Reissner-Nordstrom AdS (RNAdS) black hole phase transition, which is another type of first order phase transition and has the similar behavior with the Van der Walls fluid \cite{Kubiznak:2012wp}, was also extensively investigated by researchers. When the transition is spatial homogeneous, the dynamics of the RNAdS black hole phase transition has been studied based on the underlying free energy landscape \cite{Li:2020nsy,Li:2021vdp}. However, the bubble nucleation process by taking the spatial inhomogeneous into account was not investigated until now. In this work, we will address the problems of the vacuum decays and the bubble nucleation in the Schwarzschild AdS (SAdS) black holes and the RNAdS black holes.

The bubble nucleation driven by the vacuum energy difference stems from Coleman's pioneer work on the false vacuum decay without gravitation \cite{Coleman:1977py,Callan:1977pt}, where the $O(4)$ symmetric Euclidean bounce solution is related to the tunneling process and the Euclidean action of the bounce solution gives the tunneling rate. Later, this approach was generalized to take the effect of gravitation into account \cite{Coleman:1980aw}. Especially, modelling the inhomogeneity by the black hole, the symmetry of the Euclidean bounce solution is reduced to $O(3)$. It is also found that the seed black hole can catalyze the bubble nucleation process \cite{Hiscock:1987hn,Gregory:2013hja,Burda:2015yfa}. In general, if a true vacuum bubble is created inside of the false vacuum, it will collapse into the seed black hole. However, this collapsing bubble can tunnel to the growing bubble that inflates to the spatial infinity or falls through the cosmological horizon. Black hole seeded false vacuum decay has very important applications in early universe \cite{Burda:2015isa,Burda:2016mou,Cai:2020ndh}, as well as in the information paradox of black holes in AdS space \cite{Chen:2018aij,Chen:2021jzx}.

In this work, we consider the false vacuum decay and the bubble nucleation in the Schwarzschild AdS black hole and the RNADS black hole. The geometry of our model is described by the spacetime with an uncharged thin wall. The interior and the exterior spacetimes separated by the bubble wall are two different AdS black holes with different cosmological constant or vacuum energy. With this setting, the driving force to generate the bubble nucleation is dominated by the difference of the vacuum energy. Using the Israel junction conditions \cite{Israel:1966rt}, we derived the equation of motion for the thin bubble wall. It is shown that for the fixed mass of the interior spacetime, there is a mass range for the exterior spacetime when the Euclidean bounce bubble solution exists. It is also shown that for the fixed mass of the exterior (initial) RNAdS black hole, there exists a minimum mass of the interior (final) RNAdS black hole that the bounce solution exists. We derived the analytical expression of the Euclidean action of the bubble wall solution. The tunneling rate can only be calculated numerically. In numerical investigations, we firstly compare our result of the tunneling coefficient from the Minkovski spacetime to AdS space with that of Coleman and De Luccia in \cite{Coleman:1980aw}. It is shown that the black hole can catalyze the bubble nucleation process. Then, we investigated the bubble nucleation rate in the SAdS black holes and the RNAdS black holes. The similar conclusion is obtained. For the bubble nucleation in the RNAdS black holes, we show that the tunneling rate to the final RNAdS black hole with the minimum critical mass is the highest among all the possible tunneling channels.

This paper is arranged as follows. In section \ref{sec:eom}, the geometry of the bubble nucleation model is introduced and the equation of motion of the bubble wall is derived by using the Israel junction conditions. In section \ref{sec:conditions}, we discuss the existence conditions of the Euclidean bounce solution in details. In section \ref{sec:action}, the Euclidean action of the bounce solution is calculated analytically. In section \ref{sec:results}, the numerical results and the corresponding discussion are presented. The conclusion is summarized in section \ref{sec:conclusion}.

\section{Thin-shell wall and equation of motion}
\label{sec:eom}

In this section, we firstly introduce the geometry of the model \cite{Hiscock:1987hn,Gregory:2013hja,Burda:2015yfa}, and then discuss the Israel junction conditions \cite{Israel:1966rt} used to determine the equation of motion of the bubble wall. The bubble wall is taken to be thin for simplicity.

The whole spacetime is separated by the thin wall into two parts. On each side of the thin wall $\mathcal{W}$, the spacetime $\mathcal{M}_{\pm}$ is described by the RNAdS metric, which has the form of  
\begin{eqnarray}
ds_{\pm}^2=-f_{\pm}(r_{\pm})dt_{\pm}^2+\frac{dr_{\pm}^2}{f_{\pm}(r_{\pm})}+r_{\pm}^2 d\Omega_{\pm}^2\;,
\end{eqnarray}
where 
\begin{eqnarray}
f_{\pm}(r_{\pm})&=&1-\frac{2M_{\pm}}{r_{\pm}}+\frac{Q_{\pm}^2}{r_{\pm}^2}+\frac{r_{\pm}^2}{L_{\pm}^2}\;,\\
d\Omega_{\pm}^2&=&d\theta^2+\sin^2\theta d\phi^2\;.
\end{eqnarray}
Here, $\pm$ denotes the exterior/interior spacetime respectively. Note that on each side of the wall, the angular coordinates $\theta$ and $\phi$ are the same. The interior spacetime is taken to be the bubble nucleation. The SAdS black hole can be obtained by setting $Q=0$, and the equations derived in the following can also be applied to the SAdS black hole.

In the present work, we consider the case that the thin shell is uncharged, and set $G=1$ without loss of generality. Therefore, we take $Q_{+}=Q_{-}=Q$. Considering the symmetry of the metrics, the bubble nucleation with $O(3)$-symmetry is described by the local coordinates on each side of the wall 
\begin{eqnarray}
X_{\pm}^a=\left\{ t_{\pm}(\lambda), r_{\pm}(\lambda), \theta, \phi \right\}\;,
\end{eqnarray}
where $\lambda$ is the proper time of the observer comoving with the wall. Because the wall is made of matter and thus the corresponding trajectory is timelike, the four velocity $u_{\pm}^a=\frac{dX_{\pm}^a}{d\lambda}$ of the wall must satisfy the normalization condition $g_{ab}u_{\pm}^a u_{\pm}^b=-1$. Thus, the normalization condition for the four velocity $u_{\pm}^a=\left(\dot{t}_{\pm},\dot{r}_{\pm},0,0\right)$ is given by the equation
\begin{eqnarray}\label{normal_con}
f_{\pm}(r_{\pm}) \dot{t}_{\pm}^2- \frac{ \dot{r}_{\pm}^2}{f_{\pm}(r_{\pm})} =1\;,
\end{eqnarray}
where the dot represents the derivative with respect to the proper time $\lambda$. We require that the thin wall or the boundary of the exterior/interior spacetime is parameterized by the trajectory equation 
\begin{eqnarray}
r_{\pm}-R(\lambda)=0\;.
\end{eqnarray}
In terms of the intrinsic coordinates $\zeta^A=(\lambda,\theta,\phi)$, the induced metric of the wall is then given by 
\begin{eqnarray}
ds^2=-d\lambda^2+R^2(\lambda)\left( d\theta^2+\sin^2\theta d\phi^2 \right)\;,
\end{eqnarray}
where the normalization condition of the four velocity of the wall is used. The above induced metric provides the first Israel junction condition \cite{Israel:1966rt}, which requires the continuity between the interior and the exterior induced metrics on the wall. It is well known that Israel junction conditions should be satisfied by the hypersurface that partitions the spacetime into two regions, provided the Einstein field equations are valid in the two regions \cite{Israel:1966rt}.

To obtain the equation of motion for the wall, one needs to consider the second Israel junction condition between the interior and the exterior spacetimes \cite{Israel:1966rt}, which is given by 
\begin{eqnarray}
[K]_{ab}=K_{ab}^{+}-K_{ab}^{-}=-8\pi G \left(S_{ab}-\frac{1}{2}h_{ab} S \right)\;,
\end{eqnarray}
where $K_{ab}^{\pm}$ is the extrinsic curvature on each side of the wall, $h_{ab}$ is the induced metric on the wall, and $S_{ab}$ is the energy momentum tensor of the wall. We assume $S_{ab}=-\sigma h_{ab}$, where $\sigma$ is the tension of the wall. 

From the trajectory equation of the wall, one can get the normal one-form as 
\begin{eqnarray}
dn^{\pm}\propto dr_{\pm}-\frac{\dot{R}}{\dot{t}_{\pm}}dt_{\pm}\;.
\end{eqnarray}
The unit vector normal to each side of the wall is given by 
\begin{eqnarray}
n_{a}^{\pm}=\left(-\dot{R}, \dot{t}_{\pm},0 ,0 \right)\;.
\end{eqnarray}
It is easy to check that the normalization of the normal vector $n_{a}^{\pm}$ can be guaranteed by Eq.(\ref{normal_con}). The extrinsic curvature is then defined as 
\begin{eqnarray}
K_{ab}^{\pm}=\nabla_{a} n_{b}^{\pm}\;.
\end{eqnarray}
One can calculate the $(\theta,\theta)$ component of the extrinsic curvature
\begin{eqnarray}
K_{\theta\theta}^{\pm}=\Gamma_{\theta\theta}^{r}n_{r}^{\pm}=R f_{\pm}(R) \dot{t}_{\pm}\;,
\end{eqnarray}
where $\Gamma_{\theta\theta}^{r}$ is the connection coefficient. Therefore, the $(\theta,\theta)$ component of the second Israel junction condition gives 
\begin{eqnarray}
f_{+}(R) \dot{t}_{+}-f_{-}(R) \dot{t}_{-}=-4\pi G\sigma R\;.
\end{eqnarray}
The $(\phi,\phi)$ component of the second Israel junction condition gives the same equation as the $(\theta,\theta)$ component. Other non-zero components (for example, $(t,t)$ and $(r,r)$ components) of the second Israel junction condition give no independent equation, but the derivatives of the above equation. Combining with the normalization condition of the unit normal vector $n_{a}^{\pm}$, one can derive the equation of motion of the wall 
\begin{eqnarray}
\dot{R}^2+U(R)&=&0\;,\nonumber\\
U(R)&=&1-\frac{2M_{+}\left(1+\Delta D\right)}{R}+\frac{Q^2}{R^2}
-\frac{D^2M_{+}^2L_{+}^2}{R^4}+\frac{\left(1-\Delta^2\right)}{L_{+}^2}R^2\;,
\end{eqnarray}
where 
\begin{eqnarray}
D&=&\frac{1}{4\pi\sigma L_{+}} \frac{\left(M_+-M_-\right)}{M_+},\nonumber\\
\Delta&=&\frac{L_+}{8\pi\sigma}\left[\frac{1}{L_-^2}-\frac{1}{L_+^2}-\left(4\pi\sigma\right)^2\right]\;. 
\end{eqnarray}
This equation determines the motion of the thin wall in real time. 

In order to calculate the Euclidean action of the bounce solution, one needs the equation of motion for the Euclidean bubble wall. By performing the Wick rotation $t_{\pm}\rightarrow-i \tau_{\pm}$ as well as the the Wick rotating $\lambda\rightarrow-i\lambda$ of the proper time of the bubble wall, one can get the equation of motion for the Euclidean bubble wall 
\begin{eqnarray}
\dot{R}^2+U_E(R)&=&0\;,\nonumber\\
U_E(R)&=&-U(R)\nonumber\\
&=&-1+\frac{2M_{+}\left(1+\Delta D\right)}{R}-\frac{Q^2}{R^2}
+\frac{D^2M_{+}^2L_{+}^2}{R^4}-\frac{\left(1-\Delta^2\right)}{L_{+}^2}R^2\;.
\end{eqnarray}
The bounce solution of this equation determines the semiclassical tunneling rate of the black hole nucleation. The details of this effective potential will be discussed in the following sections.

\section{Existence conditions of the bounce solution}
\label{sec:conditions}

In our set up, there are six parameters, that is $\left\{M_+,M_-,Q,L_+,L_-,\sigma\right\}$. Recall that $M_{\pm}$ and $L_{\pm}$ are the mass and the cosmological constant of the exterior/interior black hole. The electric charge of the spacetime is set to $Q$, and $\sigma$ is the tension of the bubble wall.

We take $\left\{L_+,L_-,\sigma\right\}$ as the input of the theoretical model. We consider the case that the vacuum energy of the interior spacetime is smaller than that of the exterior spacetime, i.e. $L_+>L_-$. In the following, we will study the possible parameter range that allows a bounce solution.

Firstly, we consider the conditions of the non-extremal black holes. Without the loss of generality, we take the exterior spacetime as illustration. If the black hole is extremal, the horizon $r_{+}^*$ must satisfy the following conditions  
\begin{eqnarray}
f_{+}(r_{+}^*)&=&1-\frac{2M_{+}}{r_{+}^*}+\frac{Q^2}{\left(r_{+}^*\right)^2}+\frac{\left(r_{+}^*\right)^2}{L_{+}^2}=0\;,\\
f_{+}'(r_{+}^*)&=&\frac{2M_{+}}{\left(r_{+}^*\right)^2}-\frac{2Q^2}{\left(r_{+}^*\right)^3}+\frac{2r_{+}^*}{L_{+}^2}=0\;.
\end{eqnarray}
Fortunately, the equations can be solved analytically, which is given by  
\begin{eqnarray}
\left(r_{+}^*\right)^2=\frac{L_{+}^2}{6}\left(\sqrt{1+\frac{12Q^2}{L_{+}^2}}-1\right)\;.
\end{eqnarray}
Substituting this expression back, one can get the critical mass of the extremal RNAdS black hole as 
\begin{eqnarray}
M_{+}^*=\frac{L_{+}}{3\sqrt{6}}\left(\sqrt{1+\frac{12Q^2}{L_{+}^2}}+2\right)\left(\sqrt{1+\frac{12Q^2}{L_{+}^2}}-1\right)^{1/2}\;.
\end{eqnarray}
When $M_+>M_+^*$, the exterior black hole has two horizons and is non-extremal. The similar condition is also applied to the interior RNAdS black hole, i.e. 
\begin{eqnarray}
M_-&>&M_-^*\;,\nonumber\\
M_{-}^*&=&\frac{L_{-}}{3\sqrt{6}}\left(\sqrt{1+\frac{12Q^2}{L_{-}^2}}+2\right)\left(\sqrt{1+\frac{12Q^2}{L_{-}^2}}-1\right)^{1/2}\;.
\end{eqnarray}
The critical mass of the extremal black hole can be shown to be the increasing function of the AdS radius. Therefore, it can be concluded that $M_+^*<M_-^*$ when $L_+>L_-$.

Now, we consider the general behaviors of the potential $U_E(R)$ in the limits of $R\rightarrow 0$ and $R\rightarrow +\infty$. The shapes of the effective potential are plotted in Sec.\ref{sec:results}. When $D\neq 0$, i.e. $M_+\neq M_-$, for $R\rightarrow 0$ limit, $U_E(R)$ is approximated by 
\begin{eqnarray}
U_E(R)\simeq \frac{D^2M_{+}^2L_{+}^2}{R^4}\;.
\end{eqnarray}
Therefore, $\left.U_E(R)\right|_{R\rightarrow 0}>0$ is always guaranteed.  

For $R\rightarrow +\infty$ limit, $U_E(R)$ can be expanded as 
\begin{eqnarray}
U_E(R)\simeq -1-\frac{\left(1-\Delta^2\right)}{L_{+}^2}R^2\;.
\end{eqnarray}
Therefore, the necessary condition for the existence of the bounce solution is $\left.U_E(R)\right|_{R\rightarrow +\infty}>0$, which gives us the condition $\Delta>1$. This condition can be further expressed as 
\begin{eqnarray}
L_-<\frac{1}{4\pi\sigma}\;,\;\;\;L_+>\frac{L_-}{1-4\pi\sigma L_-}\;.
\end{eqnarray}
Under these conditions, it can be shown that for the fixed interior black hole mass $M_-$, there exists a parameter range of $M_+$, in which the potential allows the bounce solution. This will be further discussed in Sec.\ref{sec:results}.

When $D=0$, i.e. $M_+= M_-$, the effective potential becomes 
\begin{eqnarray}
U_E(R=-1+\frac{2M_{+}}{R}-\frac{Q^2}{R^2}
-\frac{\left(1-\Delta^2\right)}{L_{+}^2}R^2\;.
\end{eqnarray}
The condition that $\Delta>1$ should also be imposed. Due to the different behavior near $R\rightarrow 0$, it can be shown that when
\begin{eqnarray}\label{Q_cond}
Q<\frac{L_+}{\sqrt{12(\Delta^2-1)}}\;,
\end{eqnarray}
there exists a critical mass $M_{+}^{**}$, which is given by 
\begin{eqnarray}
M_{+}^*=\frac{L_{+}}{3\sqrt{6}\left(\Delta^2-1\right)^{1/2}}\left(\sqrt{1+\frac{12\left(\Delta^2-1\right)Q^2}{L_{+}^2}}+2\right)\left(1-\sqrt{1+\frac{12\left(\Delta^2-1\right)Q^2}{L_{+}^2}}\right)^{1/2}\;. 
\end{eqnarray}
When $M_+>M_{+}^{**}$, the effective potential has three intersection points, i.e. there exists bounce solution. Comparing this critical mass with the critical mass of the extreme black hole, one can conclude that $M_{+}^{**}$ is always smaller than $M_{+}^{*}$. This implies that when $D=0$, i.e. $M_+= M_-$, there always exists a bounce solution provided Eq.(\ref{Q_cond}) is satisfied.

In summary, the necessary conditions for the existence of Euclidean bounce solution are (1) the interior/exterior RNAdS black hole is non-extremal; (2) $\Delta>1$. For the special case of $D=0$, The existence of the bounce solution restricts the parameter range of the electric charge $Q$.

\section{Euclidean action and decay rate around RNAdS black hole}
\label{sec:action}

In this section, we derive the semiclassical tunneling rate of bubble that nucleated in the RNAdS black hole. In general, if a Lorentzian bubble is created inside of the false vacuum, it will collapse into the original black hole. However, considering the quantum effects, this collapsing bubble can tunnel to the growing bubble that inflates to the AdS spatial infinity. In the last section, we have discussed the necessary conditions that the Euclidean bounce solution exists. It is well known that the Euclidean bounce solution oscillates between the tunneling points of the corresponding Lorentzian bubble solution and the semiclassical tunneling rate of Lorentzian bubble is related to the Euclidean action of the bounce solution \cite{Coleman:1977py,Callan:1977pt,Coleman:1980aw}.

Now, we calculate the Euclidean action of the bounce solution. We start with Einstein-Hilbert action \cite{Gibbons:1976ue}
\begin{eqnarray}
I_E=&-&\frac{1}{16\pi} \int_{\mathcal{M_+}}\left( \mathcal{R}_{+}-2 \Lambda_+\right)\sqrt{g} d^4x-\frac{1}{16\pi} \int_{\mathcal{M_-}}\left( \mathcal{R}_{-}-2 \Lambda_-\right)\sqrt{g} d^4x\nonumber\\
&-&\frac{1}{16\pi} \int_{\mathcal{M_+\cup M_-}}F_{ab}F^{ab}\sqrt{g} d^4x + \int_{\mathcal{W}} \sigma \sqrt{h} d^3x\nonumber\\
&+&\frac{1}{8\pi} \int_{\partial M_+} K^+ \sqrt{h} d^3x
-\frac{1}{8\pi} \int_{\partial M_-} K^- \sqrt{h} d^3x\;,
\end{eqnarray}
It is obvious that the trace of the energy-momentum tensor for the electromagnetic field is zero. For the RNAdS solution, we have $\mathcal{R}_{\pm}=4\Lambda_{\pm}$ in the exterior and interior spacetimes. The Israel junction condition at the thin wall gives $K^+-K^-=-12\pi G \sigma$. Note that for the AdS spacetime, we have $\Lambda_{\pm}=-\frac{3}{L_{\pm}^2}$. The Euclidean time $\tau_{\pm}$ in the action integral has the period $\beta$, which is selected to be the period of the bounce solution rather than the period $\beta_-$ that can remove the canonical singularity at the horizon of the interior spacetime.

For the interior spacetime $\mathcal{M}_-$, the gravitational bulk term contains the extra contribution from the canonical singularity \cite{Gregory:2013hja,Burda:2015yfa}. It can be shown that this contribution is proportional to the horizon area times the deficit angle of the canonical singularity \cite{Fursaev:1995ef,Solodukhin:2011gn,Nishioka:2018khk}. Therefore, we have
\begin{eqnarray}
I_{\mathcal{M}_-}&=&-\frac{1}{16\pi} \int_{\mathcal{M_-}}\left( \mathcal{R}_{-}-2 \Lambda_-\right)\sqrt{g} d^4x\nonumber\\
&=&-\frac{1}{4}\left(1-\frac{\beta}{\beta_-}\right) \mathcal{A}_{-}-\frac{1}{4} \int  \frac{2}{3}\Lambda_-\left(R^3-r_h^3\right)d\tau_-\;\;,
\end{eqnarray}
where $\beta_-$ and $\mathcal{A}_-$ are the inverse temperature and the horizon area of the interior black hole, respectively.
By taking the differential of the metric function $f_-(R)$ and using the relation among the mass, the area and the temperature of the RNAdS black hole, one can get  
\begin{eqnarray}
\frac{2}{3}\Lambda_- R^3=2M_--\frac{2Q^2}{R}-R^2 f_-'\;,\nonumber\\
\frac{\mathcal{A}_-}{\beta_-}+\frac{2}{3}\Lambda_- r_h^3 -2M_-+\frac{2Q^2}{r_h}=0,\nonumber
\end{eqnarray}
Using the the two equations, we can get 
\begin{eqnarray}
I_{\mathcal{M}_-}&=&-\frac{\mathcal{A}_{-}}{4}+\frac{\beta}{4}\left[
\frac{\mathcal{A}_-}{\beta_-}+\frac{2}{3}\Lambda_- r_h^3 -2M_-
\right] +\frac{1}{4}\int \left[R^2f_{-}'+\frac{2Q^2}{R}\right]\dot{\tau}_{-}d\lambda\;\nonumber\\&=&-\frac{\mathcal{A}_{-}}{4} -\frac{\beta Q^2}{2r_h} +\frac{1}{4}\int \left[R^2f_{-}'+\frac{2Q^2}{R}\right]\dot{\tau}_{-}d\lambda\;.
\end{eqnarray}
The analytical derivation can only reach this result and the numerical method should be invoked to deal with the rest.

For the exterior spacetime $\mathcal{M}_+$, we have to add a cutoff boundary $r_0$, and subtract the divergent volume contribution from the pure AdS space. However, one has to require that the metrics of the two geometries at the cutoff surface are the same. The trick is to require that the time-periodicity in the counter term agrees with that at the cutoff boundary \cite{Witten:1998qj,deHaro:2000vlm}
\begin{eqnarray}
\beta_0=\beta\frac{f_+^{1/2}}{\left(1-\Lambda_+ r_0^2/3\right)^{1/2}}
\simeq \left(1+\frac{3M_+}{\Lambda_+ r_0^3}\right)\beta\;.
\end{eqnarray}
The action integral of the exterior spacetime with the counterterm is given by 
\begin{eqnarray}
I_{\mathcal{M}_+}&=&-\frac{1}{4} \int \frac{2}{3} \Lambda_+\left(r_0^3-R^3\right)d\tau_+ +\frac{1}{4}\int \frac{2}{3} \Lambda_+ r_0^3 d\tau_0
\;\nonumber\\
&=&\frac{1}{2}\beta M_+ +\frac{1}{4}\int \frac{2}{3} \Lambda_+ R^3 d\tau
\;\nonumber\\
&=&
\beta M_+ -\frac{1}{4}\int \left[R^2f_{+}'+\frac{2Q^2}{R}\right]\dot{\tau}_{+}d\lambda\;,
\end{eqnarray}
where in the last step, we use the fact 
\begin{eqnarray}
\frac{2}{3}\Lambda_+ R^3=2M_+-\frac{2Q^2}{R}-R^2 f_+'\;.\nonumber
\end{eqnarray}

For the contribution of the electromagnetic field, using the fact that the gauge potential is given by $A=\frac{Q}{r} dt_{\pm}$, one can directly calculate the action integral as \cite{Gibbons:1976ue}
\begin{eqnarray}
I_{EM}=\frac{\beta Q^2}{2r_h}\;.
\end{eqnarray}

For the thin wall $\mathcal{W}$, using the Israel junction conditions, we have 
\begin{eqnarray}
I_{\mathcal{W}}&=&\int_{\mathcal{W}} \sigma \sqrt{h} d^3x+\frac{1}{8\pi} \int_{\mathcal{W}}\left( K^+ - K^-\right) \sqrt{h} d^3x\nonumber\\
&=&-\frac{1}{2}\int_{\mathcal{W}} \sigma \sqrt{h} d^3x\nonumber\\
&=&\frac{1}{2} \int \left(R f_+ \dot{\tau}_+ -R f_- \dot{\tau}_-\right)d\lambda\;. 
\end{eqnarray}

At last, we have to calculate Euclidean action of the background RNAdS black hole, which gives \cite{Chamblin:1999hg,Caldarelli:1999xj}
\begin{eqnarray}
I_B=-\frac{\mathcal{A}_+}{4}+\beta M_+\;,
\end{eqnarray}
where $\mathcal{A}_+$ is the horizon area of the exterior sapcetime. Note that the integral is also performed on the period of the bounce solution and the contribution from the canonical singularity has been taken into account.

By combining the previous results, we can get the tunneling coefficient $B=I_E-I_B$ as \footnote{In principle, one does not have to subtract the action of the background spacetime. The background part is just a constant and not relevant to the bubble solution.}
\begin{eqnarray}
B&=&I_{\mathcal{M}_+}+I_{\mathcal{M}_-}+I_{\mathcal{W}}+I_{EM}-I_B\nonumber\\
&=&\frac{1}{4} \left(\mathcal{A}_+- \mathcal{A}_-\right) 
+\frac{1}{2}\int\left[\left(R-3M_+\right)\dot{\tau}_+
-\left(R-3M_-\right)\dot{\tau}_-\right]d\lambda\;.
\end{eqnarray}
This result indicates that the tunneling coefficient $B$ depends on the bounce solution, which is only solved numerically. Although the integral is performed on one period of the bounce solution, this result does not depend on the period $\beta$ explicitly.

The semiclassical tunneling rate is then given by \cite{Coleman:1977py,Callan:1977pt}
\begin{eqnarray}\label{Gamma}
\Gamma\propto e^{-B}\;. 
\end{eqnarray}
This is the semi-classical tunneling rate that the bubble is nucleated in a false vacuum of the RNAdS black hole spacetime. It can also be applied to calculate the tunneling rate of the bubble nucleation in the SAdS black holes.

\section{Numerical results}
\label{sec:results}

\subsection{Bubble nucleation in Schwarzschild black holes}

In this subsection, we consider the false vacuum decay from the Schwarzschild black hole in asymptotically flat space to the AdS space. It is shown that the black hole can catalyze the bubble nucleation process from the flat Minkovski spacetime to AdS space.

\begin{figure}
  \centering
  \includegraphics[width=8cm]{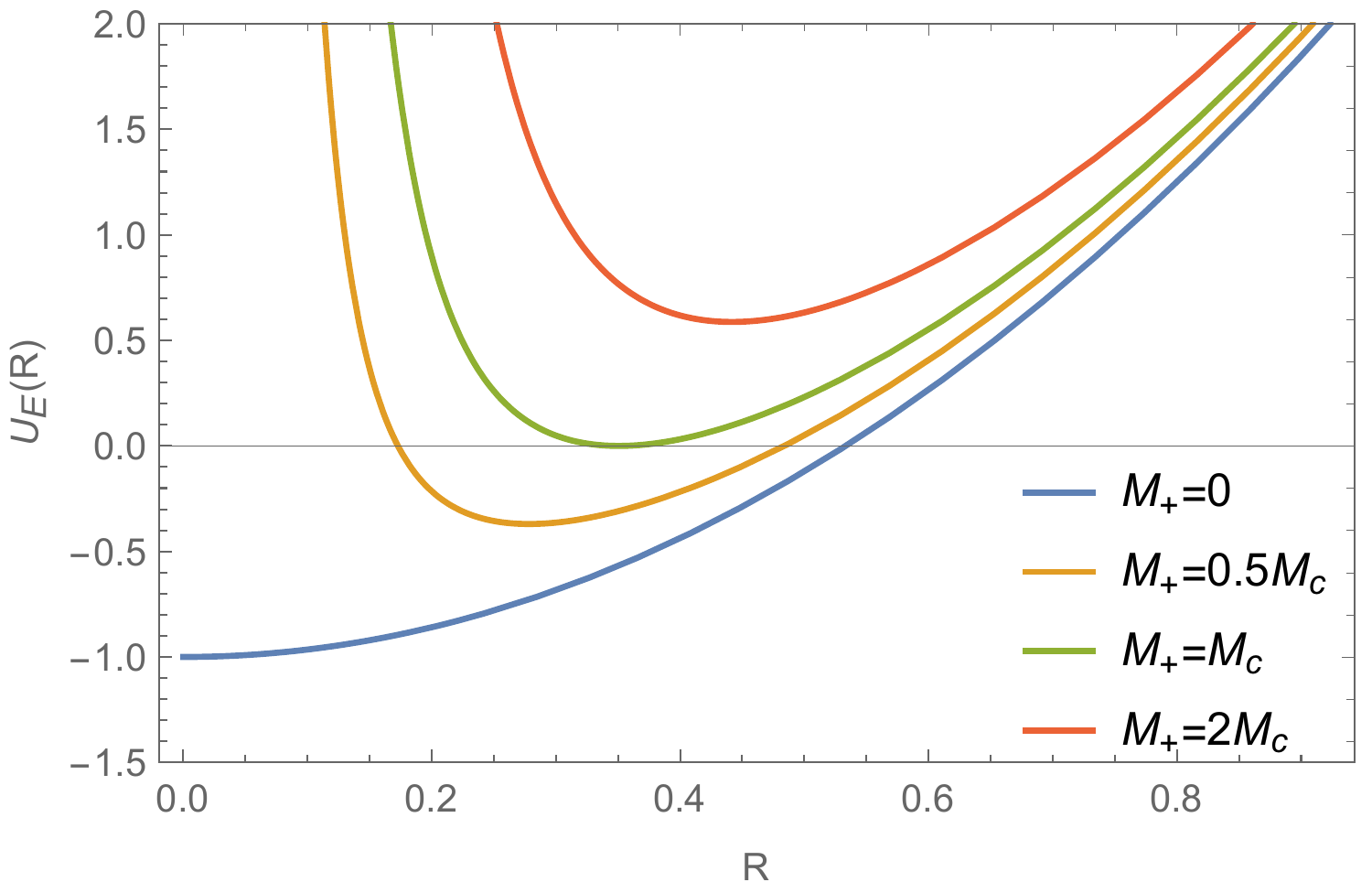}\\
  \caption{The effective potential $U_E(R)$ of the bubble nucleation in Schwarzschild black hole for different black hole mass $M_+$. In this plot, $G=1$, $L_-=1$, and $\sigma=\frac{1}{16\pi G}$. The critical black hole mass is $M_c=0.00967$. When $0<M_+<M_c$, the effective potential has two intersections with the $R$ axial. When $M_+>M_c$, there is no intersection. }
  \label{U_R_Plot_MtoAdS}
\end{figure}

By taking $Q=0$, $M_-=0$, and $L_+\rightarrow +\infty$, the effective potential can be reduced to 
\begin{eqnarray}
U_E(R)=1-\frac{2M_+}{r}-\left[\frac{r}{8\pi G\sigma}\left(\frac{1}{L_-^2}-\left(4\pi G\sigma\right)^2\right) +\frac{M_+}{4\pi G \sigma r^2}\right]^2\;.
\end{eqnarray}
In this case, the existence condition for the Euclidean bounce solution discussed in section \ref{sec:conditions} is reduced to 
\begin{eqnarray}
4\pi \sigma L_-<1\;.
\end{eqnarray}
In Figure \ref{U_R_Plot_MtoAdS}, we have plotted the effective potential for the different mass of the seed black hole. There is a critical black hole mass $M_c$, below which the bounce solution exists. The critical mass $M_c$ can be obtained by simultaneously solving the equation $U_E(R)=U_E'(R)=0$. When $0<M_+<M_c$, the bounce solution can be solved numerically, which in turn can be used to calculate the tunneling coefficient $B$.

In order to compare with the Coleman and De Luccia's (CDL) case, we quote the CDL's result on the tunneling coefficient $B$, which is given by 
\begin{eqnarray}
B_{CDL}=\frac{\pi L_-^2}{G}\frac{(4\pi G \sigma L_-)^4}{(1-(4\pi G \sigma L_-)^2)^2}\;.
\end{eqnarray}
This equation gives the tunneling coefficient from the flat Minkovski spacetime to the AdS space, which corresponds to our case with $M_+=0$.

\begin{figure}
  \centering
  \includegraphics[width=8cm]{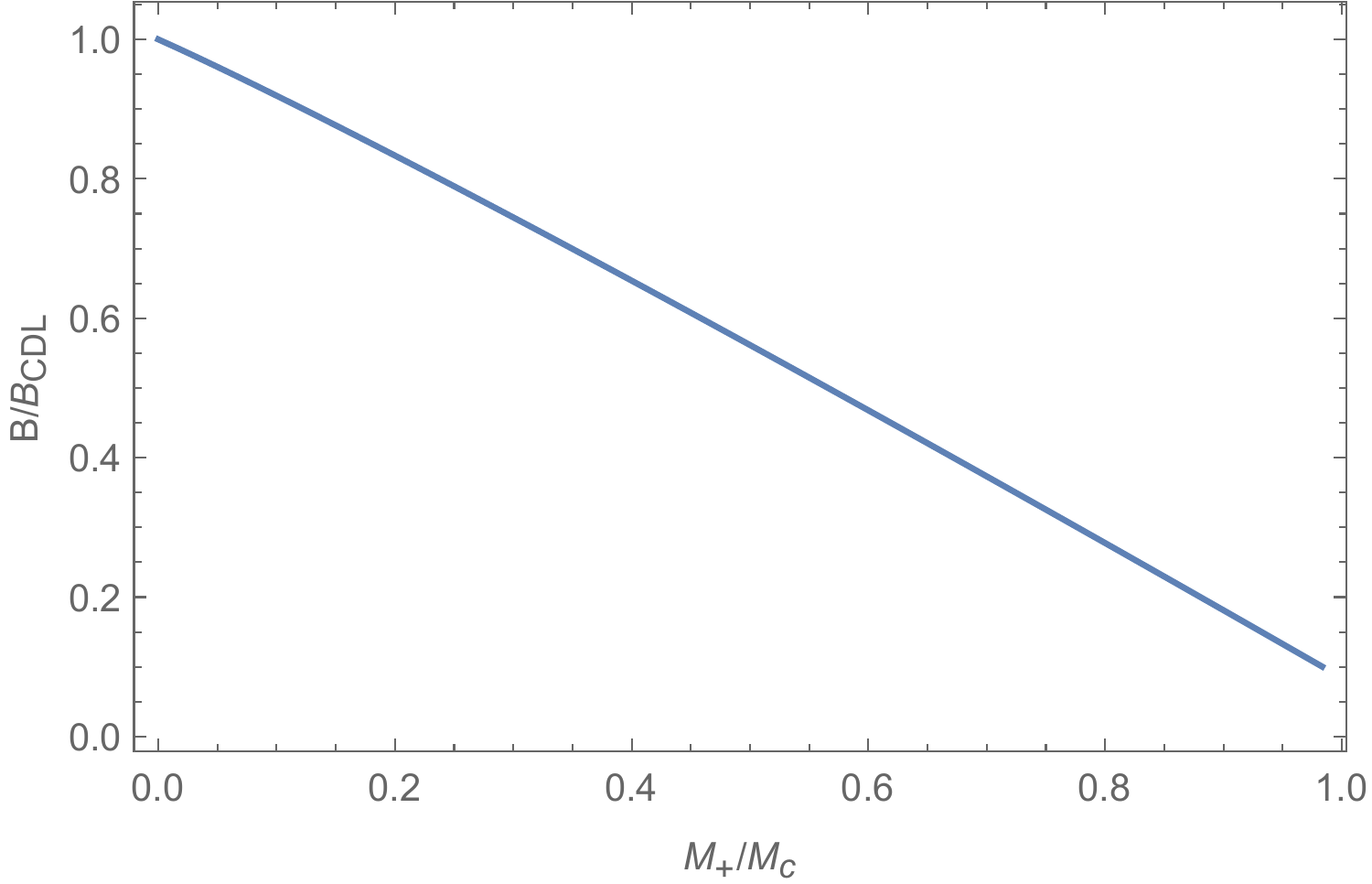}\\
  \caption{The ratio of the tunneling coefficient $B$ with $B_{CDL}$ of the bubble nucleation in Schwarzschild black hole as the function of $M_+/M_c$. In this plot, $G=1$, $L_-=1$, $\sigma=\frac{1}{16\pi G}$, and $M_c=0.00967$.}
  \label{B_M_Plot_MtoAdS}
\end{figure}

In Figure \ref{B_M_Plot_MtoAdS}, we present the numerical results on the tunneling coefficient $B$. The behaviour of the tunneling coefficient $B/B_{CDL}$ along with the mass $M_+$ of the seed black hole is explicitly plotted. Our numerical result perfectly reproduces the CDL result at the limit $M_+\rightarrow 0$. It is also shown that when increasing the black hole mass $M_+$, the tunneling coefficient $B$ decreases. According to 
Eq.(\ref{Gamma}), the the tunneling rate $\Gamma$ is then the increasing function of the black hole mass $M_+$. Therefore, the black hole can catalyze the bubble nucleation process from the flat Minkovski spacetime to the AdS space.

\subsection{Bubble nucleation in SAdS black holes}

In this subsection, we consider the transition from the SAdS black hole to the AdS space towards the bubble nucleation. The nucleation process can be viewed as the Hawking-Page phase transition from the Schwarzschild AdS black hole to the AdS space. The relation between the bubble nucleation and information paradox was previously studied in \cite{Sasaki:2014spa,Chen:2017suz,Chen:2018aij,Chen:2021jzx}.

By taking $M_-=0$ and $Q=0$, the effective potential is reduced to 
\begin{eqnarray}
U_E(R)=-1+\frac{2M_{+}\left(1+\Delta D\right)}{R}
+\frac{D^2M_{+}^2L_{+}^2}{R^4}-\frac{\left(1-\Delta^2\right)}{L_{+}^2}R^2\;,
\end{eqnarray}
with 
\begin{eqnarray}
D=\frac{1}{4\pi\sigma L_+}\;.
\end{eqnarray}
The existence condition of the bounce solution in this case is $\Delta>1$, which is consistent with the condition given in \cite{Sasaki:2014spa,Chen:2017suz}. However, the tunneling coefficient was not explicitly calculated in \cite{Sasaki:2014spa,Chen:2017suz}. We now present the numerical results of the bubble nucleation rate in the SAdS black holes.

\begin{figure}
  \centering
  \includegraphics[width=8cm]{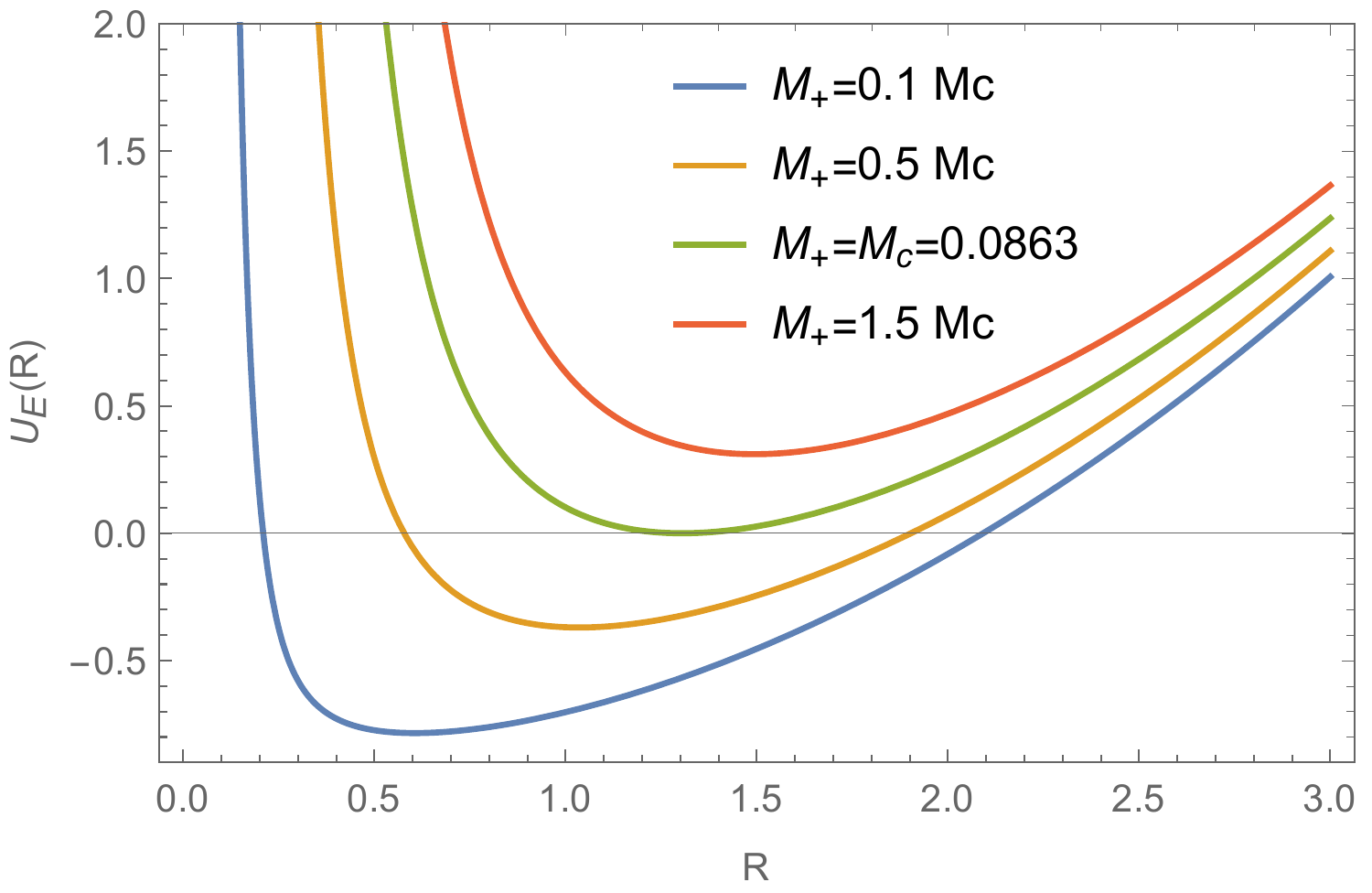}\\
  \caption{The effective potential $U_E(R)$ of the bubble nucleation in SAdS black hole for different black hole mass $M_+$. In this plot, $G=1$, $L_+=1.4$, $L_-=1$, and $\sigma=\frac{1}{16\pi G}$. The critical black hole mass is $M_c=0.0863$. When $0<M_+<M_c$, the effective potential has two intersections with the $R$ axial. When $M_+>M_c$, there is no intersection. }
  \label{U_R_Plot_HP}
\end{figure}

The effective potential of the bubble nucleation in SAdS black hole is plotted in Figure \ref{U_R_Plot_HP}. It is shown that there is a critical mass $M_c$ of the initial black hole. When $M_+=M_c$, the potential curve is tangent to the $R$-axial. When $0<M_+<M_c$, there exists the Euclidean bounce solution. When the mass of the initial SAdS black hole is in this range, the phase transition to the AdS space can occur towards the bubble nucleation process.

\begin{figure}
  \centering
  \includegraphics[width=8cm]{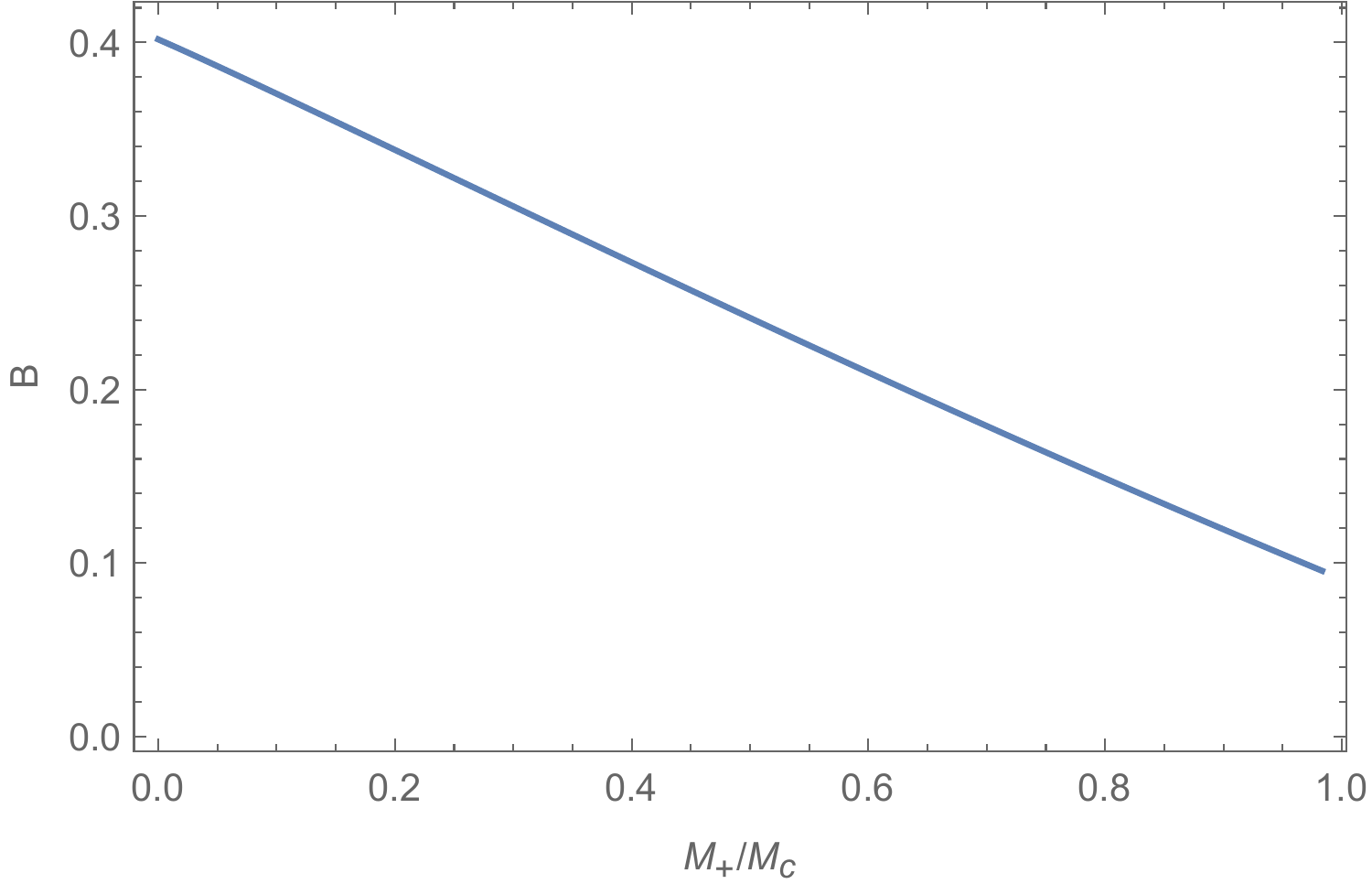}\\
  \caption{The tunneling coefficient $B$ of the bubble nucleation in SAdS black hole as the function of $M_+/M_c$. In this plot, $G=1$, $L_+=1.4$, $L_-=1$, $\sigma=\frac{1}{16\pi G}$, and $M_c=0.0863$.}
  \label{B_M_Plot_HP}
\end{figure}

In Figure \ref{B_M_Plot_HP}, the numerical results for the tunneling coefficient are presented. It is shown that when the initial black hole mass increases, the tunneling rate increases also. In this case, we can also conclude that the black hole as the seed of the nucleation can catalyze the transition process. The bubble nucleation process from the SAdS black hole to the AdS space is the analogy of the Hawking-Page phase transition. However, the Euclidean bounce solution only exists when the seed black hole mass is smaller than the critical mass $M_c$. This implies that the transition via the bubble nucleation process occurs only for the SAdS black hole with the small mass.

\subsection{Bubble nucleation in RNAdS black holes}

In this subsection, we consider the transition from one RNAdS black hole to another RNAdS black hole via the bubble nucleation. This process is analogy to the RNAdS black hole phase transition. However, the two RNAdS black holes have different AdS radius. From the extended phase space viewpoint \cite{Kubiznak:2012wp}, this means that the two RNAdS black holes have different thermodynamic pressures.

\begin{figure}
  \centering
  \includegraphics[width=8cm]{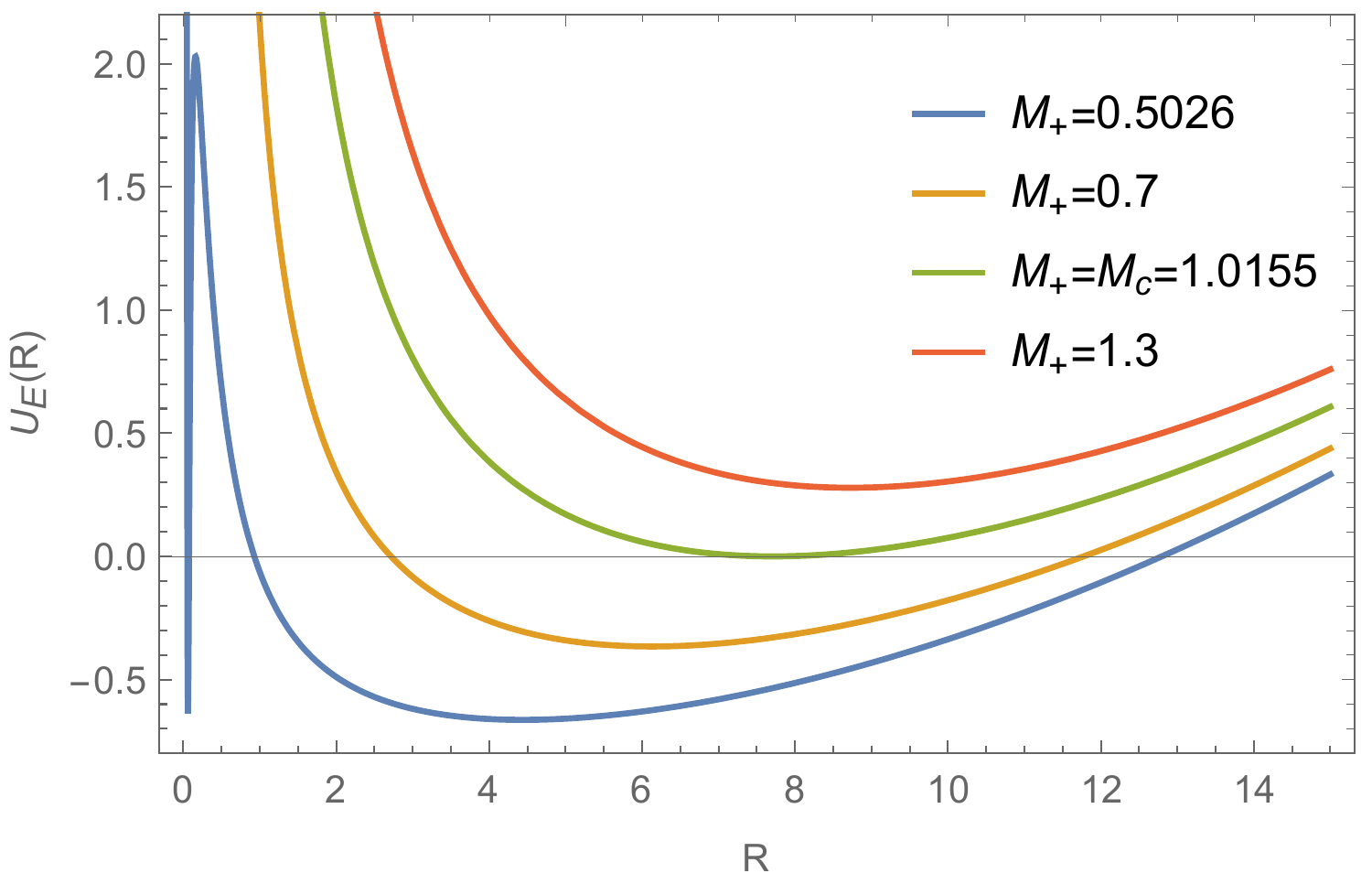}\\
  \caption{The effective potential $U_E(R)$ with the fixed mass $M_-$ of the final black hole. In this plot, $G=1$, $L_+=1.335$, $L_-=1$, $\sigma=\frac{1}{16\pi G}$, $Q=0.3$, and $M_-=0.5$. The critical mass of the initial black hole is $M_c=1.0155$. }
  \label{U_R_Plot_RNAdS}
\end{figure}

In Figure \ref{U_R_Plot_RNAdS}, the effective potentials of the bounce solutions in the RNAdS black holes are plotted for different masses $M_+$ of the initial black holes and fixed final black hole mass $M_-$. The behavior is similar to the potential discussed in the previous subsections. There exists a critical mass for the initial black hole, beyond which the bounce solution doesn't exist. Because of the existence of the term $-\frac{Q^2}{R^2}$ in the effective potential, the plot of the potential for $M_+=0.5026$ has a sharp barrier in the small $R$ region. However, the small $R$ behavior is irrelevant to the bounce solution that governs the tunneling rate.

\begin{figure}
  \centering
  \includegraphics[width=8cm]{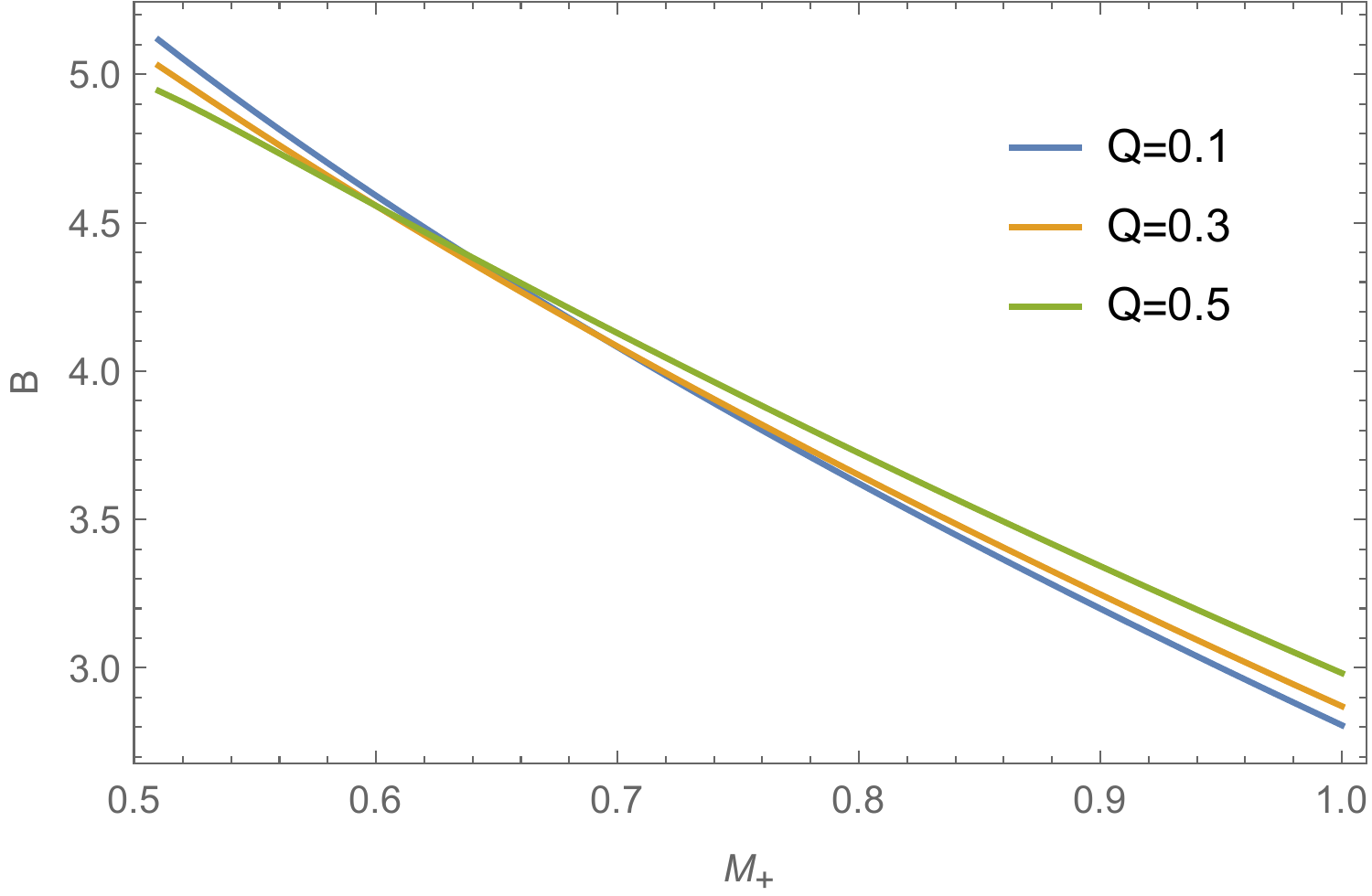}\\
  \caption{The tunneling coefficient $B$ is plotted as the function of the initial black hole mass $M_+$. In this plot, $G=1$, $L_+=1.335$, $L_-=1$, $\sigma=\frac{1}{16\pi G}$, $Q=0.3$, and $M_-=0.5$. The critical mass of the initial black hole is $M_c=1.0155$.}
  \label{B_M_Plot_RNAdS}
\end{figure}

In Figure \ref{B_M_Plot_RNAdS}, we present the numerical results for the relationship between the tunneling coefficient $B$ and the initial black hole mass $M_+$. It is shown that increasing the mass $M_+$ will decrease the tunneling coefficient $B$, which in turn will increase the tunneling rate of the bounce solution. The charged black hole can also catalyze the bubble nucleation process in the AdS space.

\begin{figure}
  \centering
  \includegraphics[width=8cm]{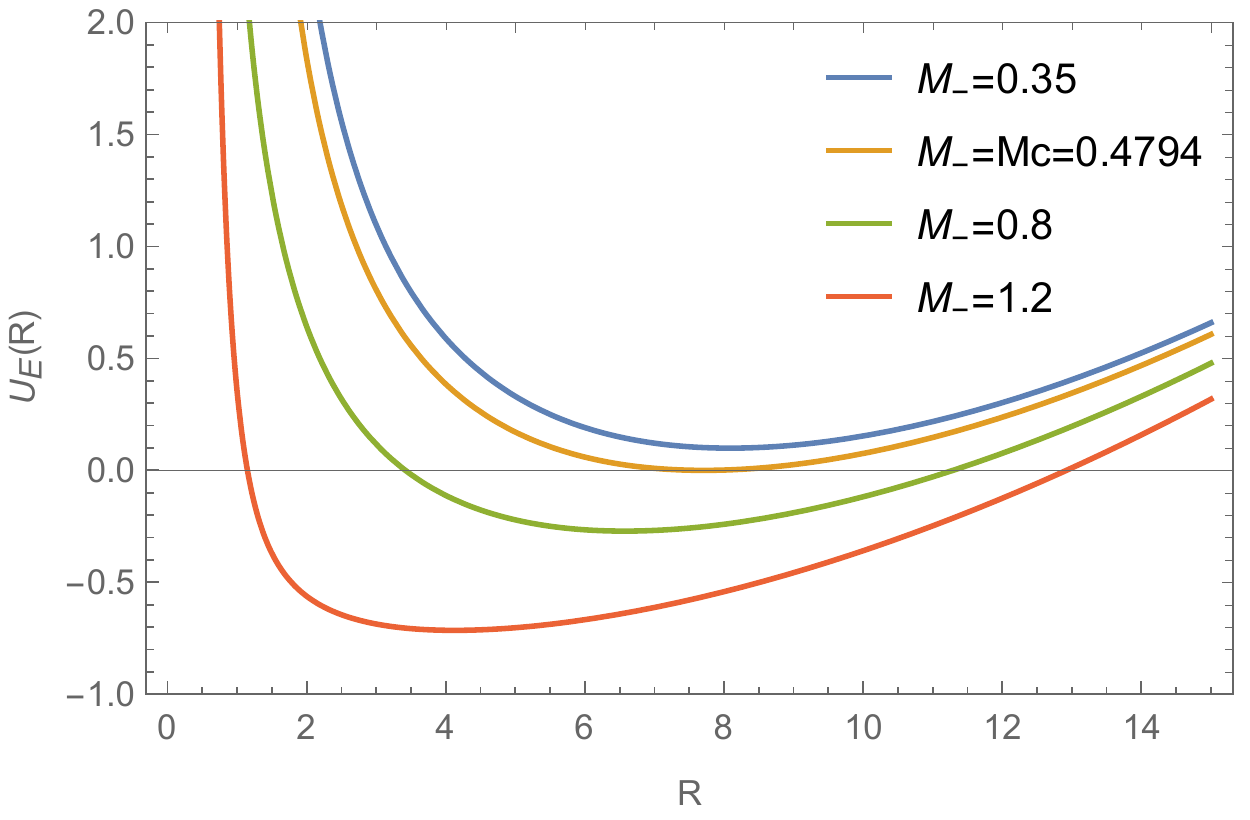}\\
  \caption{The effective potential for the fixed initial RNAdS black hole mass. In this plot,  $G=1$, $L_+=1.335$, $L_-=1$, $\sigma=\frac{1}{16\pi G}$, $Q=0.3$, and $M_+=1$. The critial mass of the final RNAdS black hole mass is $M_c=0.4794$.  }
  \label{U_R_Plot_RNAdS_diffMn}
\end{figure}

In Figure \ref{U_R_Plot_RNAdS_diffMn}, we consider the case that the initial black hole mass is kept fixed while the final black hole mass can vary. The plots shows that there also exists a critical mass $M_c$ for the final black hole mass. This critical mass is the minimum mass for the final black hole that the initial black hole can decay to. There is no other constraint on the final black hole mass.

\begin{figure}
  \centering
  \includegraphics[width=8cm]{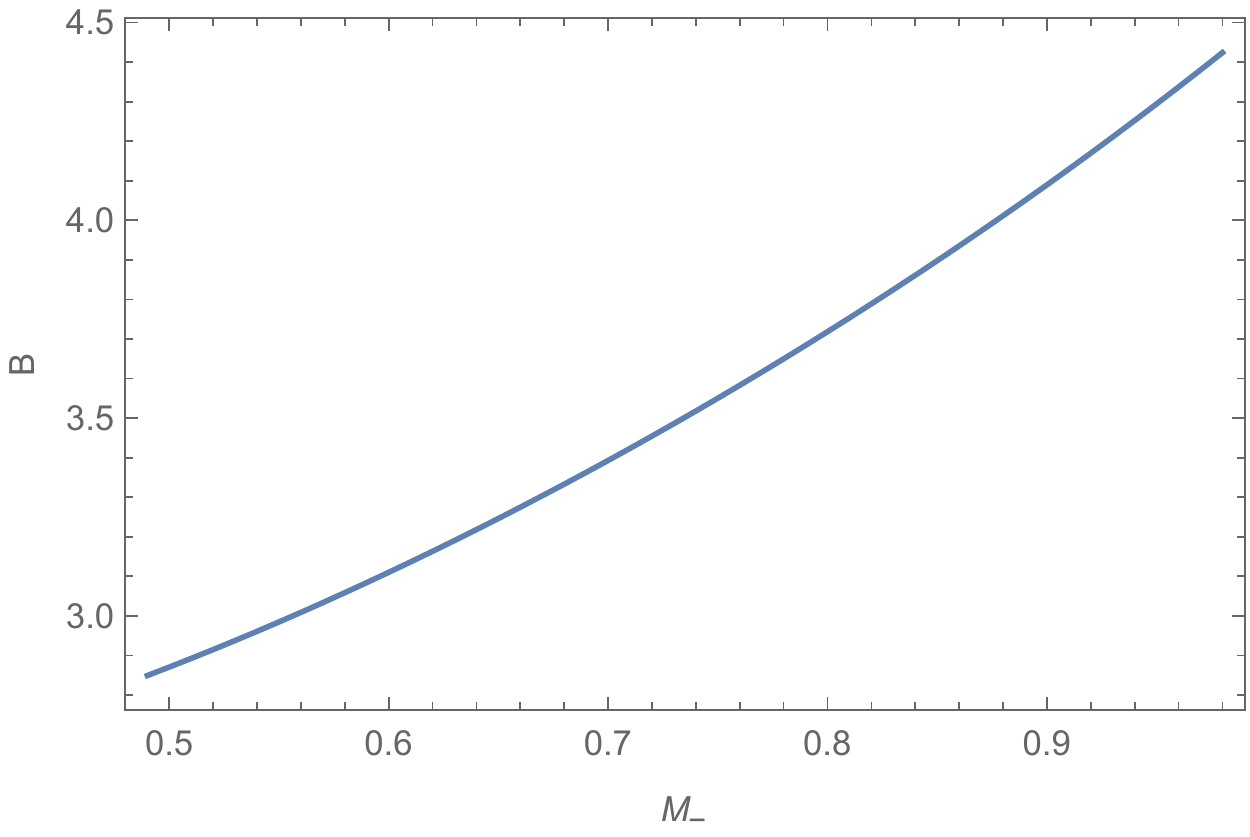}\\
  \caption{The tunneling coefficient $B$ verses the final RNAdS black hole mass $M_-$. The range of $M_-$ is from the minimum critical mass to a relative large value. In this plot, $G=1$, $L_+=1.335$, $L_-=1$, $\sigma=\frac{1}{16\pi G}$, $Q=0.3$, and $M_+=1$. The critical mass of the final RNAdS black hole mass is $M_c=0.4794$. }
  \label{B_Mn_Plot_RNAdS}
\end{figure}

In Figure \ref{B_Mn_Plot_RNAdS}, we plot the tunneling coefficient $B$ for the fixed initial black hole. Note that the range of $M_-$ is from the minimum critical mass to a relative large value. It is shown that the tunneling coefficient is the increasing function of the final black hole mass. This implies that it is harder for the initial black hole to decay to the final black hole with the bigger mass.
As shown in Figure \ref{U_R_Plot_RNAdS_diffMn}, when the initial black hole mass $M_+$ is fixed, there is a minimum critical mass $M_c$ for the final black hole that allows a bounce solution. The plot in Figure \ref{B_Mn_Plot_RNAdS} shows that the tunneling rate to the final RNAdS black hole with the minimum critical mass is the highest. On the other hand, due to the fact that there is no bounce solution below the minimum critical mass, Figure \ref{U_R_Plot_RNAdS_diffMn} and Figure \ref{B_Mn_Plot_RNAdS} also indicate that the initial RNAdS black hole can not decay to the final RNAdS black hole with arbitrary small mass. This result shows the tunneling channel from the initial fixed RNAdS black hole to the final RNAdS black hole.

\section{Conclusion}
\label{sec:conclusion}

In this work, we have studied the false vacuum decay and the bubble nucleation in the Schwarzschild AdS black holes and the RNADS black holes. Starting from the geometric setting of the bubble wall spacetime and using the Israel junction conditions, we derived the equation of motion for the bubble wall. For all the cases that we have considered, if the mass of the black hole inside the bubble wall is fixed, there is a mass range for the exterior black hole spacetime that the Euclidean bounce bubble solution exists. It is also shown that for the fixed mass of the exterior (initial) RNAdS black hole, there exists a minimum mass of the interior (final) RNAdS black hole that the bounce solution exists.

We then derived an analytical expression of the Euclidean action for the bounce solution. However, the tunneling coefficient can only be calculated numerically. For the numerical results, we firstly compare our result of the tunneling coefficient from the Minkovski spacetime to AdS space with that of Coleman and De Luccia. It is found that the numerical result perfectly reproduce the CDL's result when the mass of the initial Schwarzschild black hole approaches zero. Then, we considered the bubble nucleation in the SAdS black hole and the RNAdS black hole. It is shown that the black hole can catalyze the bubble nucleation process. In particular, for the bubble nucleation in the RNAdS black holes, we show that the tunneling rate to the final RNAdS black hole with the minimum critical mass is the highest among all the possible tunneling channels.

At last, we should point our the thin wall model considered in the present work is uncharged. In general, the bubble wall produced in the charged spacetime background should be charged also \cite{Kuchar:1968,Lemos:2021jtm}. For the future direction, it is interesting to consider the charged thin wall that separates the interior and the exterior spacetimes. Another interesting aspect is to study the effect of inhomogeneity by the black hole on vacuum decay and bubble nucleation in higher derivative gravity and modified gravity theories \cite{Cai:2008ht,Charmousis:2008ce,Salehian:2018yoq}.


\end{document}